\documentclass[a4paper,10pt]{article}
\usepackage[margin=0.9in]{geometry}
\usepackage{microtype}
\usepackage[utf8x]{inputenc}
\usepackage[english]{babel}
\usepackage[babel]{csquotes}
\usepackage{authblk}
\usepackage{mathrsfs}
\usepackage{indentfirst}
\usepackage{amssymb,amsmath}
\usepackage{array}
\usepackage{epsfig}
\usepackage{graphicx}
\usepackage{latexsym}
\usepackage{array}
\usepackage{fancyhdr}
\usepackage{pstricks}
\usepackage{bm}
\usepackage{slashed}
\usepackage{color}
\usepackage{booktabs}
\usepackage{braket}

\usepackage[normalem]{ulem}  

\newcommand{\beq}{\begin{equation}}
\newcommand{\eeq}{\end{equation}}
\newcommand{\eq}{Eq.~}
\newcommand{\eqs}{Eqs.~}
\newcommand{\fig}{Fig.~}
\newcommand{\reff}{Ref.~}
\renewcommand{\Re}{\operatorname{Re}}

\renewcommand\sout{\bgroup \color{blue} \ULdepth=-.5ex \ULset}

\begin{document}

\title{Polarization and 
  dilepton anisotropy in\\ pion-nucleon collisions}
\author{Enrico Speranza$^{1,2}$, Mikl\'os Z\'et\'enyi$^{3,4,5}$, and Bengt Friman$^1$ \\
\emph{\small $^1$GSI Helmholtzzentrum f\"ur Schwerionenforschung GmbH, D-64291 Darmstadt, Germany}\\
\emph{\small $^2$Technische Universit{\"a}t Darmstadt, D-64289 Darmstadt, Germany}\\
\emph{\small $^3$Wigner Research Center for Physics, H-1121 Budapest, Hungary}\\
\emph{\small $^4$Extreme Matter Institute EMMI, GSI, D-64291 Darmstadt, Germany}\\
\emph{{\small $^5$Institut f\"ur Theoretische Physik, Goethe-Universit{\"a}t, D-60438 Frankfurt am Main, Germany}}\\
}

\date{}

\maketitle 

\begin{abstract}
  Hadronic polarization and the related anisotropy of the dilepton angular 
  distribution are studied for the reaction $\pi N \rightarrow Ne^+ e^-$. 
  We employ consistent effective interactions for baryon resonances up to 
  spin-5/2, where non-physical degrees of freedom  are eliminated, 
  to compute the anisotropy coefficients for isolated intermediate baryon
  resonances. It is shown that the spin and parity
  of the intermediate baryon resonance is reflected in the 
  angular dependence of the anisotropy coefficient. We then compute the
  anisotropy coefficient including the $N(1520)$ and $N(1440)$ resonances,
  which are essential at the collision energy of
  the recent data obtained by the HADES collaboration on this reaction. 
  We conclude that  the anisotropy coefficient provides useful 
  constraints for unravelling the resonance contributions to this process.
\end{abstract}

\section{Introduction}
Dilepton production in hadronic reactions provides information on the 
electromagnetic properties of hadrons. Leptons are also  important 
probes of nuclear collisions, since their mean-free path in nuclear matter is much larger than nuclear sizes. Hence, they can carry information on the conditions that prevail during the brief, highly compressed stages of 
the reaction. Dileptons are produced in a variety of different elementary 
processes.  Multiply differential cross sections for dilepton production can provide 
information needed to disentangle the production channels.

Independently of the specific reaction, dileptons originate from the 
decay of virtual photons. According to the vector meson dominance hypothesis, 
hadrons couple to the electromagnetic field via a neutral vector meson, which 
subsequently converts into a photon. Although vector meson dominance does not provide
an accurate description of the electromagnetic coupling for all hadrons, it
provides a tenable first approximation. This implies that dilepton 
production in nuclear collisions can furnish information on the in-medium spectral 
functions of vector mesons.

A detailed understanding of elementary hadronic reactions is an important 
prerequisite for studies of nuclear collisions. While a lot of effort has 
been invested in the study of dilepton production in nucleon-nucleon collisions both 
in experiment and theory, pion-nucleon collisions are less explored. 
The HADES collaboration has recently studied pion-induced reactions, including 
dilepton production. First preliminary
data have been presented at the NSTAR2015 conference \cite{HADES}.
The aim of the present paper is to explore the reaction 
$\pi N \rightarrow R \rightarrow Ne^+e^-$, where $R$ is the intermediate baryon resonance, 
in terms of effective Lagrangian models 
at the center-of-momentum (CM) energy of the HADES experiment. In 
particular we study the angular distribution of the produced dileptons.

The general expression for the angular distribution of dileptons
originating from the decay of a virtual photon is given
by~\cite{Gottfried,Falciano,Faccioli}:
\begin{equation}
  \label{eq:angdist}
  \frac{d \sigma}{d \Omega_e}\propto 1+\lambda_\theta \cos^2 \theta_e 
  + \lambda_{\theta\phi} \sin 2\theta_e \cos \phi_e 
  + \lambda_\phi \sin^2 \theta_e \cos 2\phi_e,
\end{equation}
where $\theta_e$ and $\phi_e$ are the polar and azimuthal angles of one
of the two leptons in the rest frame of the photon. The anisotropy coefficients $\lambda_\theta$, $\lambda_{\theta\phi}$ and $\lambda_{\phi}$ depend on the choice of the quantization axis. We use the so-called helicity frame where the polarization
axis is chosen along the momentum of the virtual photon in the CM frame \cite{HERA}. 

As we discuss in Section
\ref{sec:xsec}, in the reaction $\pi N \rightarrow R \rightarrow Ne^+e^-$, the anisotropy coefficients depend on the quantum numbers of the intermediate baryon resonance and on the scattering angle $\theta_{\gamma^*}$ of the virtual photon. In \reff \cite{Bratkovskaya:1995kh} the coefficient
$\lambda_\theta$ has been computed for some of the relevant dilepton
sources in heavy-ion collisions. On the experimental side, a non-zero polarization of the $J/\psi$ has been found in proton-proton
\cite{LHCb} and proton-nucleus collisions \cite{HERA}. In heavy-ion collisions the anisotropy coefficients is compatible with zero at low invariant masses of the lepton pair~\cite{NA60}.

In pion-nucleon scattering, 
it is expected that a major part of the dilepton production cross 
section is due to $s$-channel baryon resonances
with a mass close to the CM energy $\sqrt{s}$. The emergence of a dilepton 
anisotropy in these processes can be understood as follows.
The initial state, which in the CM frame contains a pion with momentum $\mathbf{p}$
and a nucleon with momentum $-\mathbf{p}$, can be expanded
in terms of eigenstates of orbital angular momentum
\begin{equation}
  \label{eq:orbital}
  \Ket{\pi(\mathbf{p});N(-\mathbf{p})} \propto 
  \sum_{lm} Y^*_{lm}(\theta,\phi) \Ket{lm},
\end{equation}
where $\theta$ and $\phi$ specify the direction of $\mathbf{p}$ with
respect to the quantization axis. Here spin quantum numbers as well as
the normalization are omitted for clarity. We choose the quantization axis $z$
parallel to the momentum of the incident pion, implying that $\theta = 0$. Since
$Y_{lm}(\theta=0,\phi) \ne 0$ only for $m=0$, the $z$-component of the
orbital angular momentum vanishes in the initial state. Hence, the projection 
of the total spin of the intermediate
baryon resonance on the beam axis is given by the $z$-component of the nucleon
spin. This means that only the $J_z=+1/2$ and $-1/2$ states of the resonance are 
populated. 

As a result, in case of an unpolarized nucleon target, spin-$1/2$ 
intermediate resonances are unpolarized, and consequently there is no 
preferred direction in the CM frame. Accordingly, in this case all
observables are independent of the scattering angle, i.e., the
angle $\theta_{\gamma^*}$ of the virtual photon in the CM frame. 
On the other hand, intermediate resonances of spin$\ge 3/2$ have a 
nontrivial polarization, implying an angular anisotropy 
in the CM frame. Consequently, in this case, observables show a nontrivial dependence
on the scattering angle $\theta_{\gamma^*}$, 
which reflects the quantum numbers of the resonance.
 
The determination of the quantum numbers of the baryon resonances 
produced in pion-nucleon collisions is important for gaining a deeper understanding 
of hadron-hadron interactions in general and dilepton production in hadronic collisions
in particular. The above arguments indicate that the study of the angular 
distribution of dileptons can provide valuable information on the 
spin and parity of the intermediate resonances. 
In principle, by comparing measured angular distributions of the dileptons
with theoretical predictions for different resonance states, it should 
be possible to set constraints on the quantum numbers of the resonance. 
The same idea has been used e.g. to determine the quantum numbers of 
the $X(3872)$ meson~\cite{Aaij:2013zoa}. 

At the intermediate energies explored in the HADES experiment,
the task is facilitated by the small number of baryon resonances that contribute
significantly. However, in practice the situation is often more complex than in the ideal case 
of an isolated resonance which dominates the cross section. Thus, in general there is interference with 
nearby resonances and with a non-resonant background which must be accounted for.      

In this paper we study the angular distribution of dileptons in the process
$\pi N \rightarrow Ne^+e^-$ in terms of the anisotropy coefficient $\lambda_\theta$ of
\eq\eqref{eq:angdist}. In
Section~\ref{sec:xsec} we give the expressions of the differential cross 
section and the anisotropy coefficient. In Section~\ref{sec:model} we
briefly review the effective Lagrangian of our model. This is followed by
a presentation of the numerical results for the anisotropy coefficient
in Section~\ref{sec:results}, where we also briefly explore the effect 
of nearby resonances. Conclusions are presented in Section~\ref{sec:summary}.

\section{\label{sec:xsec} Cross section and anisotropy coefficient}
The differential cross section of the process $\pi N \rightarrow
Ne^+e^-$ can be written in the form
\begin{equation}
  \label{eq:dsdm}
  \frac{d\sigma}{dM d\cos\theta_{\gamma^*} d\Omega_{e}} = 
  \frac{M}{64(2\pi)^4 s}\frac{|\mathbf{p}_f|}{|\mathbf{p}_i|}  
  \frac{1}{n_\text{pol}} \sum_\text{pol}
  \left\vert\mathcal{M}\right\vert^2,
\end{equation}
where $\mathcal{M}$ is the matrix element, $\theta_{\gamma^*}$ is the polar angle of the momentum of the
virtual photon in the CM frame measured from the beam axis and 
$d\Omega_{e}$ is the solid angle of the electron in the rest frame of the 
lepton pair \cite{Zetenyi:2012hg}. Moreover, $M$ is the invariant mass of the lepton pair, $s$ is the 
square of the CM energy,  $\mathbf{p}_i$ and 
$\mathbf{p}_f$ are the CM three-momenta of the initial and final nucleons, 
respectively. The sum is over all spin states in the initial 
and final state particles and $n_\text{pol}$ is the number of spin polarizations 
in the initial state.

The square of the matrix element can be written in the form
\begin{equation}\label{eq:msquared}
  \sum_\text{pol} \left\vert\mathcal{M}\right\vert^2 = \frac{e^2}{k^4}
  W_{\mu\nu}l^{\mu\nu},
\end{equation}
where the lepton tensor,
\begin{equation}
  l^{\mu\nu} = 4\left[k_1^{\mu}k_2^{\nu} + k_1^{\nu}k_2^{\mu} -
  (k_1\cdot k_2+m_e^2)g^{\mu\nu} \right]
\end{equation}
describes the coupling of the virtual photon to the $e^+e^-$ pair,
$k_1$ and $k_2$ being the momenta of the electron and positron, respectively, and $m_e$ being the electron mass.
The quantity
\begin{equation}
  W_{\mu\nu} = \sum_\text{pol}
  \mathcal{M}_{\mu}^{\text{had}}{\mathcal{M}_{\nu}^{\text{had}}}^{*}
\end{equation}
is the hadronic tensor, where $\mathcal{M}_{\mu}^{\text{had}}$ is the
hadronic part of the matrix element $\mathcal{M}$.

The nontrivial angular distribution of the $e^+e^-$ pair is connected with
the polarization state of the decaying virtual photon. Therefore it is
convenient to work in the polarization density matrix
representation \cite{Choi:1989yf}.
Let $\epsilon^{\mu}(k,\lambda)$ denote the polarization
vector of the virtual photon of momentum $k$ (the helicity $\lambda$ can 
take on values $\pm$1 and 0,
since virtual photons can also be longitudinally polarized). The polarization vectors for the three helicities are, in the rest frame of the virtual photon,
\begin{align}
\epsilon^{\mu}(k,-1)&=\frac{1}{\sqrt{2}}(0,1,-i,0), \\
\epsilon^{\mu}(k,0)&=(0,0,0,1), \\
\epsilon^{\mu}(k,+1)&=-\frac{1}{\sqrt{2}}(0,1,i,0). 
\end{align}
We define the hadronic (or production) density matrix as
\begin{equation}
  \rho^{\text{had}}_{\lambda,\lambda^{\prime}} = \frac{e^2}{k^4} 
  \epsilon^{\mu}(k,\lambda) W_{\mu\nu} \epsilon^{\nu}(k,\lambda^{\prime})^*
\end{equation}
and the leptonic (or decay) density matrix as
\begin{equation}
  \rho^{\text{lep}}_{\lambda^{\prime},\lambda} = 
  \epsilon^{\mu}(k,\lambda^{\prime}) l_{\mu\nu} 
  \epsilon^{\nu}(k,\lambda)^*.
\end{equation}
In terms of the density matrices the square of the matrix element is given by
\begin{equation}
  \label{eq:M2}
  \sum_\text{pol} \left\vert\mathcal{M}\right\vert^2 = 
    \sum_{\lambda,\lambda^{\prime}}
  \rho^{\text{had}}_{\lambda,\lambda^{\prime}} 
    \rho^{\text{lep}}_{\lambda^{\prime},\lambda}.
\end{equation}
The equivalence of \eqs\eqref{eq:msquared} and \eqref{eq:M2}
is seen by employing the identity
\begin{equation}
  \sum_{\lambda} \epsilon^{\mu}(k,\lambda) \epsilon^{\nu}(k,\lambda)^* =
  - g^{\mu\nu} + \frac{k^{\mu}k^{\nu}}{k^2}
\end{equation}
and the fact that $k^{\mu} l_{\mu\nu} = 0$.

The explicit form of the lepton spin density matrix
$\rho^{\text{lep}}_{\lambda^{\prime},\lambda}$ is 
given by
\begin{equation}
  \label{eq:rho_lep}
  \rho^{\text{lep}}_{\lambda^{\prime},\lambda} = 4|{\bf k}_1|^2 \left(
  \begin{array}{ccc}
    1+\cos^2\theta_e+\alpha & -\sqrt{2}\cos\theta_e\sin\theta_e e^{-i\phi_e} & 
      \sin^2\theta_e e^{-2i\phi_e} \\
    -\sqrt{2}\cos\theta_e\sin\theta_e e^{i\phi_e} & 2(1-\cos^2\theta_e)+\alpha &
      \sqrt{2}\cos\theta_e\sin\theta_e e^{-i\phi_e} \\
    \sin^2\theta_e e^{2i\phi_e} & 
      \sqrt{2}\cos\theta_e\sin\theta_e e^{i\phi_e} & 1+\cos^2\theta_e +\alpha
  \end{array}
  \right),
\end{equation}
where $\theta_e$ and $\phi_e$ are the polar and azimuthal angle of one
of the two lepton momenta relative to the polarization axis in the rest
frame of the virtual photon, ${\bf k}_1$ is the three-momentum of one of the two leptons in the virtual photon rest frame and $\alpha=\frac{2m_e^2}{|{\bf k}_1|^2}$. As noted above, we use the so-called helicity frame. In the rest of the paper we neglect the electron mass. 
The angular dependence of the squared matrix element is obtained by 
combining Eqs.\ \eqref{eq:M2} and \eqref{eq:rho_lep},
\begin{align}
	\label{eq:msq_spin-dens}
  \sum_\text{pol} \left\vert\mathcal{M}\right\vert^2 & \propto
  (1+\cos^2\theta_e )(\rho^{\text{had}}_{-1,-1} + \rho^{\text{had}}_{1,1})
  + 2(1-\cos^2\theta_e )\rho^{\text{had}}_{0,0} \nonumber\\
    & + \sqrt{2}\cos\theta_e\sin\theta_e
  \left[e^{i\phi_e}( \rho^{\text{had}}_{0,1}-\rho^{\text{had}}_{-1,0} )
    + e^{-i\phi_e}(\rho^{\text{had}}_{1,0} - \rho^{\text{had}}_{0,-1})
  \right] \nonumber \\
  & + \sin^2\theta_e (e^{2i\phi_e}\rho^{\text{had}}_{-1,1} 
  + e^{-2i\phi_e}\rho^{\text{had}}_{1,-1}).
\end{align}
Here we suppressed the explicit dependence of $\rho^{\text{had}}_{\lambda,\lambda^{\prime}}$ on $M$ and $\theta_{\gamma^*}$.
By comparing Eqs.~(\ref{eq:angdist}) and (\ref{eq:msq_spin-dens}), we can identify
the anisotropy coefficients
\begin{align}
  \lambda_\theta & = \frac{\rho^{\text{had}}_{-1,-1}+\rho^{\text{had}}_{1,1}-2\rho^{\text{had}}_{0,0}}
  {\rho^{\text{had}}_{-1,-1}+\rho^{\text{had}}_{1,1}+2\rho^{\text{had}}_{0,0}}, \\
    \lambda_{\theta\phi} & = \sqrt{2}
  \frac{\Re(\rho^{\text{had}}_{0,1}-\rho^{\text{had}}_{-1,0})}
  {\rho^{\text{had}}_{-1,-1}+\rho^{\text{had}}_{1,1}+2\rho^{\text{had}}_{0,0}}, \\
  \lambda_\phi & = 2 \frac{\Re(\rho^{\text{had}}_{-1,1})}                                               
  {\rho^{\text{had}}_{-1,-1}+\rho^{\text{had}}_{1,1}+2\rho^{\text{had}}_{0,0}}, 
\end{align}
where we used the fact that the density matrix is hermitian.  
Equation \eqref{eq:msq_spin-dens} contains two further terms, proportional to $\sin 2\theta_e \sin\phi_e$ and $\sin^2 \theta_e \sin 2\phi_e$, respectively, which in principle define two additional anisotropy coefficients. However, these two additional terms are unobservable \cite{Faccioli}.

The interpretation of the $\lambda_\theta$ coefficient becomes clear if we 
integrate over the azimuthal angle of the electron momentum $\phi_e$.
In this case only the first two terms of \eq\eqref{eq:msq_spin-dens} yield a nonzero result,
and hence only the diagonal elements of
$\rho^{\text{had}}_{\lambda,\lambda^{\prime}}$ contribute to the 
averaged cross section. Consequently, the angular distribution \eq\eqref{eq:dsdm} can be
cast in the form
\begin{equation}
  \label{eq:dsdm2}
  \frac{d\sigma}{dM d\cos\theta_{\gamma^*} d\cos \theta_{e}}  
  \propto \Sigma_\bot (1+ \cos^2 \theta_e) + \Sigma_\parallel 
  (1-\cos^2 \theta_e) ,
\end{equation}
where $\Sigma_\bot=\rho^{\text{had}}_{-1,-1} +
\rho^{\text{had}}_{1,1}$ and $\Sigma_\parallel=2\rho^{\text{had}}_{0,0}$
are the contributions of the transverse and parallel polarizations of the
intermediate photon to the differential cross section.
Equation \eqref{eq:dsdm2} 
can be rearranged in the following way: 
\beq 
\label{eq:dsdm3}
\frac{d\sigma}{dM d\cos\theta_{\gamma^*} d\cos
  \theta_{e}} \propto \mathcal{N}(1+\lambda_\theta\cos^2\theta_e), 
\eeq
where the anisotropy coefficient $\lambda_\theta$ is given by
\beq
  \label{eq:bcoeff}
  \lambda_\theta (M, \theta_{\gamma^*})=\frac{\Sigma_\bot - \Sigma_\parallel}{\Sigma_\bot + \Sigma_\parallel}.
\eeq 
Thus,
the anisotropy coefficient provides information on the polarization
of the virtual photon. If the virtual photon is created via the decay of an intermediate
baryon resonance, then $\lambda_\theta$ (and in particular its dependence on 
$\theta_{\gamma^*}$) depends on the quantum numbers of the baryon resonance.

Predictions for the $\lambda_\theta$ coefficient for some dilepton sources are given
in \reff\cite{Bratkovskaya:1995kh}. In particular for the
Drell-Yan process $\lambda_\theta=+1$ (the virtual photon is completely
transversely polarized), while for the pion annihilation process $\lambda_\theta=-1$
(the virtual photon is completely longitudinally polarized). In the case
of a medium in thermal equilibrium, the polarization of the virtual photon,
and hence the anisotropy coefficient $\lambda_\theta$, in general depends on the 
momentum relative to the heat bath in addition to the production mechanism \cite{Hoyer:1986pp}.

\section{\label{sec:model} The model}
In the present study we first focus on the contribution of Feynman diagrams
containing one baryon resonance in the $s$-channel, and set the resonance mass equal 
to $\sqrt{s}$. Thus, we can explore the 
dependence of the angular shape of the anisotropy coefficient on the quantum numbers of the 
intermediate state in the ideal case of an isolated on-shell resonance. We then include 
the relevant baryon resonances in the mass range between 1.4 and 1.68 GeV and add also the 
corresponding $u$-channel diagrams. Obviously, this choice of diagrams is not 
exhaustive. Nevertheless, it provides a first estimate of interference effects.
For the interpretation of the HADES data at $\sqrt{s} = 1.49$~GeV, the nearby 
resonances $N(1440)$, $N(1520)$ and $N(1535)$ are expected to provide the leading
contributions.

We assume that baryons couple to the electromagnetic field via an
intermediate $\rho^0$ meson according to the vector meson dominance
model.\footnote{We neglect the isoscalar channel, i.e. the $\omega$ meson,
since it is not expected to modify the polarization of the virtual photon. We note,
however, that $\rho-\omega$ interference~\cite{Lutz:2002nx} may affect the dependence of the 
polarization observables on the dilepton invariant mass.} For the $\rho^0$-photon coupling we use the gauge invariant vector meson dominance 
model Lagrangian~\cite{Kroll:1967it}
\begin{equation}
  \label{eq:VMD}
  \mathcal{L}_{\rho\gamma} = - \frac{e}{2g_{\rho}} F^{\mu\nu}
  \rho^0_{\mu\nu},
\end{equation}
where $F^{\mu\nu} = \partial_{\mu}A_{\nu} - \partial_{\nu}A_{\mu}$ is
the electromagnetic field strength tensor and $\rho^0_{\mu\nu} =
\partial_{\mu}\rho^0_{\nu} - \partial_{\nu}\rho^0_{\mu}$.

We include baryon resonances up to spin-5/2. The interaction of spin-1/2 baryons with 
pions and $\rho$ mesons is described by the Lagrangian densities of 
Ref.~\cite{Zetenyi:2012hg},
\begin{eqnarray}
  \mathcal{L}_{R_{1/2}N\pi} & = & - \frac{g_{RN\pi}}{m_{\pi}} 
  \bar{\psi}_R \Gamma
  \gamma^{\mu}\vec{\tau}\psi_N \cdot \partial_{\mu}\vec{\pi}
  + \text{h.c.}, \label{eq:RNpi_1h} \\
   \mathcal{L}_{R_{1/2}N\rho} & = & \frac{g_{RN\rho}}{2m_{\rho}} 
  \bar{\psi}_{R}
  \vec{\tau} \sigma^{\mu\nu} \tilde{\Gamma} \psi_N \cdot \vec{\rho}_{\mu\nu} +
  \text{h.c.}. \label{eq:RNrho_1h}
\end{eqnarray}
Here, and also in the Lagrangians involving higher spin resonances given below,
$\Gamma = \gamma_5$ for $J^P = 1/2^+$, $3/2^-$ and $5/2^+$
resonances and $\Gamma = 1$ otherwise, and $\tilde{\Gamma} =
\gamma_5\Gamma$.

Higher spin fermions are represented by Rarita-Schwinger spinor fields in
effective Lagrangian models. These fields transform according to a product of a spin-1/2
and one or more spin-1 representations of the Lorentz group. Therefore they contain
some contributions describing the propagation of lower-spin states.
In a consistent Lagrangian involving higher-spin baryons, the  
lower-spin components of the Rarita-Schwinger fields should not contribute to observable
quantities.

Such a consistent interaction scheme for spin-3/2 fermions was developed by
Pascalutsa \cite{Pascalutsa:1998pw} and generalized for spin-5/2 fermions by
Vrancx et al.\ \cite{Vrancx:2011qv}. In this work we specify the Lagrangian
describing the interaction of higher-spin baryon resonances based on the scheme of 
Ref.\ \cite{Vrancx:2011qv}. In this scheme, the lower-spin degrees of 
freedom are eliminated from observables by requiring that the Lagrangian is invariant
under the gauge transformations
\begin{eqnarray}
\label{eq:spinorgauge1}
  \psi_{\mu} & \rightarrow & \psi_{\mu} + i\partial_{\mu}\chi, \label{eq:gauge_3h}\\
  \label{eq:spinorgauge2}
  \psi_{\mu\nu} &\rightarrow & \psi_{\mu\nu} + \frac{i}{2}(\partial_{\mu}\chi_{\nu} + \partial_{\nu}\chi_{\mu}), \label{eq:gauge_5h}
\end{eqnarray}
for spin-3/2 ($\psi_{\mu}$) and spin-5/2 ($\psi_{\mu\nu}$) Rarita-Schwinger fields, respectively. In the above equations, $\chi$ and $\chi_\mu$
are arbitrary spinor and spinor-vector fields, respectively. The gauge invariance of Eqs.\ \eqref{eq:gauge_3h}
and \eqref{eq:gauge_5h} is ensured if only the gauge invariant expressions of the fields
\begin{eqnarray}
  G_{\mu,\nu} & = & i(\partial_{\mu}\psi_{\nu} - \partial_{\nu}\psi_{\mu}), \\
  G_{\mu\nu,\lambda\rho} & = & 
  - \partial_{\mu}\partial_{\nu}\psi_{\lambda\rho} - \partial_{\lambda}\partial_{\rho}\psi_{\mu\nu}
  + \frac{1}{2}(\partial_{\mu}\partial_{\lambda}\psi_{\nu\rho} + \partial_{\mu}\partial_{\rho}\psi_{\nu\lambda}
  + \partial_{\nu}\partial_{\lambda}\psi_{\mu\rho} + \partial_{\nu}\partial_{\rho}\psi_{\mu\lambda})
\end{eqnarray}
appear in the Lagrangian. Furthermore, by defining the fields
\begin{eqnarray}
  \Psi_{\mu} & = & \gamma^{\nu}G_{\mu,\nu}, \nonumber \\
  \Psi_{\mu\nu} & = & \gamma^{\lambda}\gamma^{\rho}G_{\mu\nu,\lambda\rho}, \label{eq:Psi_field}
\end{eqnarray}
and making the replacements
\begin{equation}
\label{eq:trans:m}
  \psi_{\mu}\rightarrow\frac{1}{m}\Psi_{\mu}, \quad\text{and}\quad \psi_{\mu\nu}\rightarrow\frac{1}{m^2}\Psi_{\mu\nu}
\end{equation}
in a ``traditional" Lagrangian containing Rarita-Schwinger fields one obtains a gauge
invariant and hence consistent Lagrangian. The mass parameter $m$ in \eq\eqref{eq:trans:m} is introduced for dimensional reasons. 

Starting from the Lagrangians of Ref.\ \cite{Vrancx:2011qv} and taking the isospin structure into account,
we obtain the expressions
\begin{eqnarray}
  \mathcal{L}_{R_{3/2}N\pi} & = & \frac{ig_{RN\pi}}{m_{\pi}^2}
  \bar{\Psi}_R^{\mu} \Gamma \vec{\tau}\psi_N\cdot\partial_{\mu}\vec{\pi}
  + \text{h.c.}, \label{eq:RNpi_3h}\\
  \mathcal{L}_{R_{5/2}N\pi} & = & -\frac{g_{RN\pi}}{m_{\pi}^4}
  \bar{\Psi}_R^{\mu\nu} \Gamma \vec{\tau}\psi_N \cdot
  \partial_{\mu}\partial_{\nu}\vec{\pi} + \text{h.c.}, \label{eq:RNpi_5h}
\end{eqnarray}
for the Lagrangians describing the resonance-nucleon-pion interaction. In the case of
$\Delta$ resonances, the Pauli matrices $\vec{\tau}$ appearing in the above and the following
Lagrangians have to be replaced by the isospin-3/2 $\rightarrow$ 1/2 transition matrices.

The resonance-nucleon-$\rho$ interaction Lagrangian is analogous to
the electromagnetic resonance-nucleon transition Lagrangian. Out of the three Lagrangians 
given in Ref.\ \cite{Vrancx:2011qv} we choose the one with the lowest number of
derivatives. After including the isospin structure of the interaction, the Lagrangians are 
given by
\begin{eqnarray}
  \mathcal{L}_{R_{3/2}N\rho} & = & \frac{ig_{RN\rho}}{4m_{\rho}^2}
  \bar{\Psi}_{R}^{\mu}\vec{\tau} \gamma^{\nu}\tilde{\Gamma} \psi_N \cdot
  \vec{\rho}_{\nu\mu} + \text{h.c.}, \label{eq:RNrho_3h}\\
  \mathcal{L}_{R_{5/2}N\rho} & = & - \frac{g_{RN\rho}}{(2m_{\rho})^4}
  \bar{\Psi}_{R}^{\mu\nu}\vec{\tau} \tilde{\Gamma}\gamma^{\rho}
  (\partial_{\mu}\psi_N) \cdot
  \vec{\rho}_{\rho\nu} + \text{h.c.}. \label{eq:RNrho_5h}
\end{eqnarray}

We explore the relevance of the different spins and parities by
computing the anisotropy coefficient with a hypothetical resonance for each spin-parity
combination, all with the same mass and width, $m_R = 1.49$~GeV and $\Gamma_R = 0.15$~GeV. 
The mass was chosen to coincide with the CM energy $\sqrt{s}$
used in our calculations, thus assuming that in the $s$-channel the resonance is  on the mass shell.

\begin{figure}[bt]
	\begin{center}
		\includegraphics{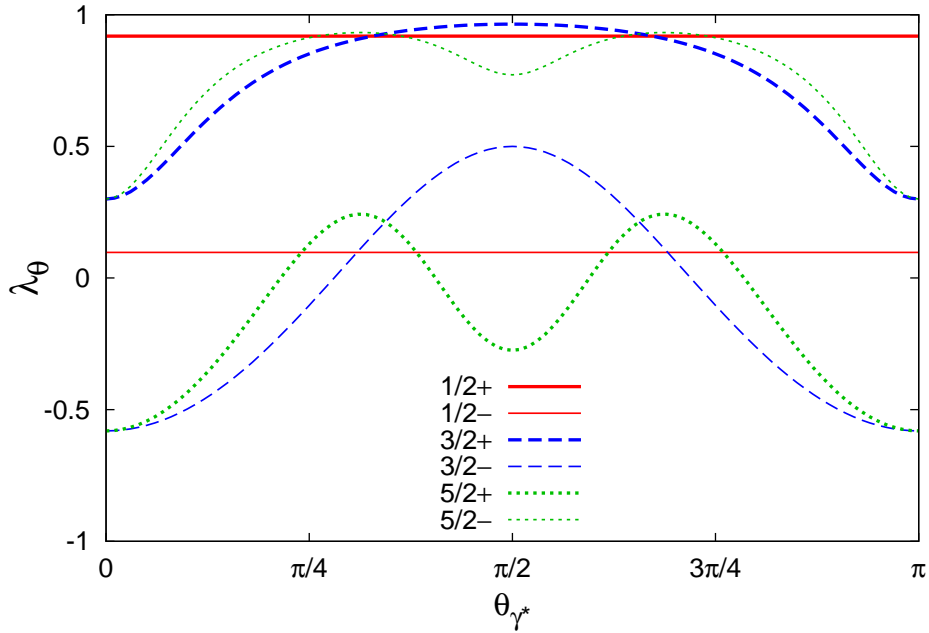}
		\caption{\label{fig:schpole}The anisotropy coefficient $\lambda_\theta$ as a function of the virtual photon polar angle $\theta_{\gamma^*}$ for hypothetical 
			resonance states with different spins and parities in the $s$-channel at a dilepton mass
			$M=0.5$~GeV. The resonance masses coincide with 
			$\sqrt{s}=1.49$~GeV, the resonance widths are $\Gamma_R = 0.15$~GeV.}
	\end{center}
\end{figure}
We also made calculations including all well established resonance
states in the relevant energy domain. These states are the nucleon
resonances $N(1440)$~$1/2^+$, $N(1520)$~$3/2^-$, $N(1535)$~$1/2^-$, $N(1650)$~$1/2^-$, 
$N(1675)$~$5/2^-$, $N(1680)$~$5/2^+$, and the $\Delta$ resonances
$\Delta(1600)$~$3/2^+$, and $\Delta(1620)$~$1/2^-$. 
The coupling constants $g_{RN\pi}$ and $g_{RN\rho}$ were determined from the widths of the
$R\rightarrow N\pi$ and $R\rightarrow N\rho\rightarrow N\pi\pi$ decays. The empirical values
for these partial widths were obtained as a product of the total width and the appropriate 
branching ratio as given by the Particle Data Group~\cite{PDG}. Masses of the resonances are also 
taken from Ref.\ \cite{PDG}.

We stress that the present model is intended to be valid for virtual photon masses not far from the $\rho$ meson mass. The gauge invariant version of the vector meson dominance \eqref{eq:VMD} does not contribute to processes with real photons and, therefore, the model has to be supplemented by a separate coupling of baryon resonances to the nucleon and photon, if we want to describe processes with low mass virtual or real photons. On the other hand, using the original (not gauge invariant) version of the vector meson dominance by Sakurai~\cite{Sakurai}, the photonic branching ratios of the baryon resonances are overpredicted \cite{Friman}. 

The $\Delta(1232)$ resonance is not included, since there is no information available on the coupling  
strengths to the $N\rho$ channel and its mass is far below the CM energy considered.

We use a simplifying approximation for the mass dependence of the
resonance width, assuming that it is given by that of the $N\pi$
channel, and employ the parametrization of Ref.~\cite{Zetenyi:2012hg}.

\section{\label{sec:results} Results}
We employ the model described above to compute the anisotropy
coefficient $\lambda_\theta$ of \eq\eqref{eq:bcoeff} for the reaction $\pi N \to N e^+
e^-$.  In the following we discuss the dependence of the anisotropy
coefficient on the polar angle of the virtual photon
$\theta_{\gamma^*}$. In all the calculations, the CM energy is set to
$\sqrt{s}=1.49$~GeV, corresponding to the HADES data.

\begin{figure}[bt]
	\begin{center}
		\includegraphics{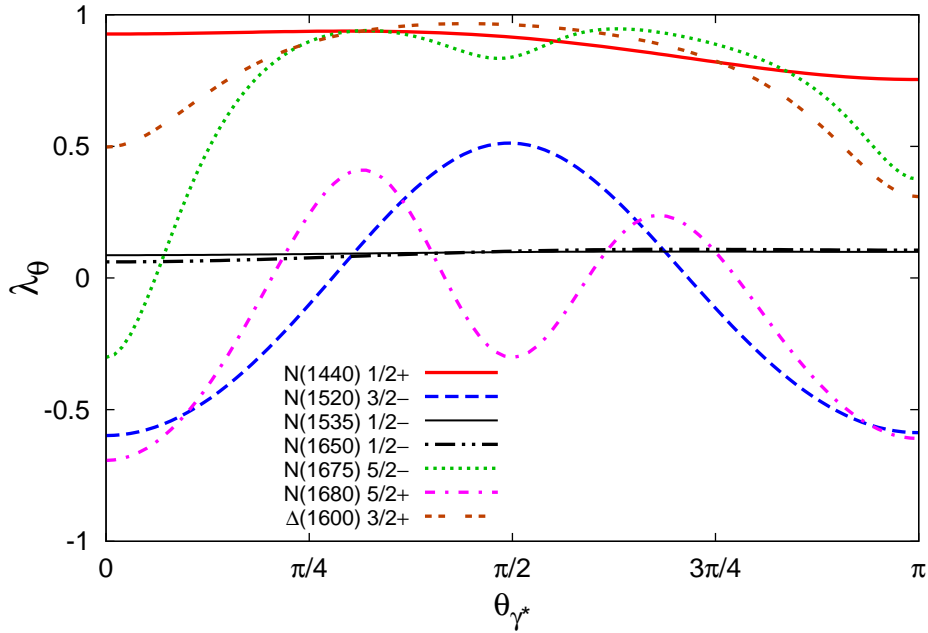}
		\caption{\label{fig:sch_uch_physical} The anisotropy coefficient $\lambda_\theta$ as a
			function of the virtual photon polar angle at a dilepton mass $M=0.5$~GeV
		including $s$- and $u$-channel diagrams. The CM energy is
			$\sqrt{s}=1.49$~GeV. For further details, see the text.}
	\end{center}
\end{figure}
In order to demonstrate the effect of different spin-parity baryon
resonance states on the $\lambda_\theta$ coefficient, we first use the model with
the hypothetical resonances discussed above, including only 
the $s$-channel Feynman diagram. In Fig.~\ref{fig:schpole}
we show the anisotropy coefficient for dileptons of invariant mass
$M=0.5$~GeV. In this case the resonance in the 
intermediate state is on-shell and, therefore, the results should
correspond to standard angular momentum coupling. Fig.~\ref{fig:schpole} shows that the spin and parity of the intermediate resonance is
reflected in a characteristic angular dependence of the anisotropy coefficient. 
In particular, in the spin-1/2
channels the $\lambda_\theta$ coefficient is independent of $\theta_{\gamma^*}$, in
accordance with the arguments given in the introduction. Based on the same arguments, the $z$-component of the total spin
coincides with the $z$-component of the initial nucleon spin. Since the
nucleon target is unpolarized, the $J_z=+1/2$ and $-1/2$
polarization states of each resonance are equally populated. This means
that there is no preferred direction and, consequently, that the
$\lambda_\theta$ coefficient is isotropic.

The dependence of the anisotropy coefficient on the quantum numbers of the intermediate baryon resonance can be  interpreted in terms of angular momentum coupling, once one accounts for the fact that, in the non-relativistic limit, the off-diagonal coupling \eq\eqref{eq:RNrho_1h} is purely transverse. We note that for all channels considered, except $J^P=1/2^+$, two values of the final state orbital angular momentum are possible. The strengths of these channels and their relative phase depend on the structure of the interaction vertices \eqs\eqref{eq:RNrho_1h}, \eqref{eq:RNrho_3h}, and \eqref{eq:RNrho_5h}. Thus, a different choice for the interaction Lagrangians~\cite{Vrancx:2011qv} may lead to a somewhat different angular dependence of the anisotropy coefficient. This indicates a certain model dependence of the results. However, as long as  the lowest angular momentum states dominate, we expect the results presented here to remain valid, at least on a qualitative level.

In \fig\ref{fig:sch_uch_physical} we show the $\lambda_\theta$ coefficient obtained
from the $s$- and $u$-channel diagrams of seven of the physical resonance
states considered. (The contribution of the $\Delta(1620)$, which has
a shape very similar to the other spin-$1/2^-$ states, is left out for the sake of clarity).
Here, the characteristic shapes presented in \fig\ref{fig:schpole} are modified mainly by the interference with the $u$-channel resonance 
contributions.

In order to assess which of the resonances are important for 
the dilepton production process at the CM energy of the HADES experiment,
we compute the differential cross section $d\sigma/dM$ by integrating 
\eq\eqref{eq:dsdm} over the scattering angle $\theta_{\gamma^*}$ and the electron 
solid angle $\Omega_{e}$. We find that at $\sqrt{s} = 1.49$~GeV and a dilepton invariant 
mass of $M=0.5$~GeV, the two dominant contributions are the $N(1520)$ with $d\sigma/dM=0.44$~$\mu$b/GeV,
and the $N(1440)$ with $d\sigma/dM=0.33$~$\mu$b/GeV. These results contain both $s$- and 
$u$-channel diagrams. The combined cross section taking
into account both $N(1520)$ and $N(1440)$ and their interference is $d\sigma/dM=0.84$~$\mu$b/GeV
when all coupling constants are assumed to have the same phase and $d\sigma/dM=0.70$~$\mu$b/GeV
if we assume that the matrix elements involving the two resonances have the opposite phase. The range of baryon resonance widths and branching ratios given by the Particle Data Group \cite{PDG} induce uncertainties in the differential cross sections. In particular, the $N(1520)$ contribution may vary by about $40\%$. For the $N\rho$ branching ratio of the $N(1440)$ resonance, only an upper limit is given. As a result, the combined total cross section including both the $N(1440)$ and $N(1520)$ can be up to a factor of $2$ larger than the values given above. On the other extreme, the $N(1440)$ branching ratio into the $\rho\,N$ channel may vanish, implying that corresponding contribution to dilepton production is negligible. Nevertheless, the average values of the branching ratios used here yield reasonable agreement with experiment, as discussed below.
\begin{figure}[bt]
	\begin{center}
		\includegraphics{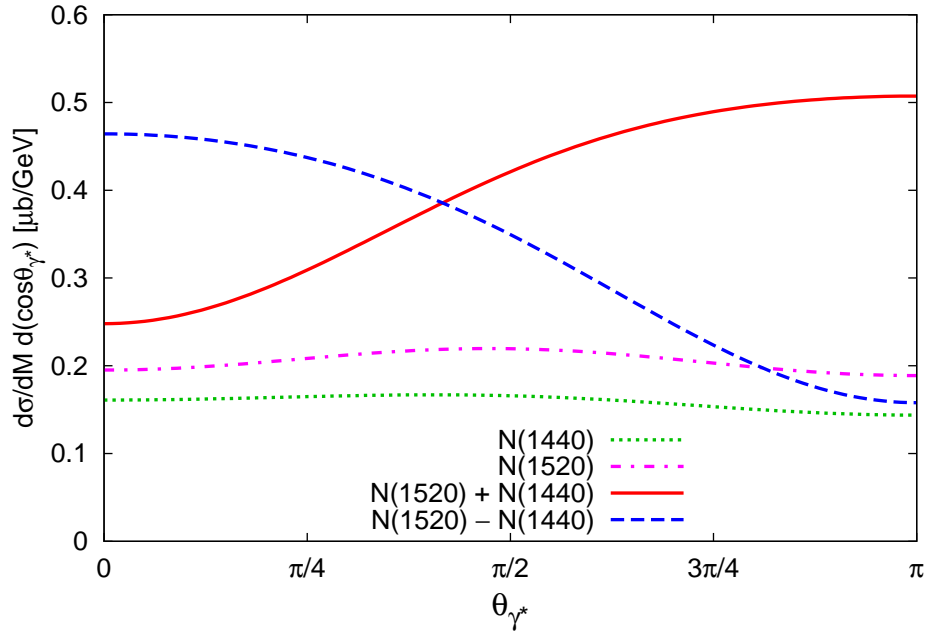}
		\caption{\label{fig:sigma_phys_dominant} The contribution of the two dominant resonances, $N(1440)$ and $N(1520)$,
			to the differential cross section of dilepton production at $\sqrt{s}=1.49$~GeV
			CM energy. Two of the curves show the result obtained from $s$- and $u$-channel
			diagrams of each resonance. The other two curves are obtained from the sum of all four diagrams
			($s$- and $u$-channel diagrams of both resonances), assuming a positive and negative relative sign
			between the amplitudes of the two resonances.}
	\end{center}
\end{figure}

The largest sub-leading contributions to the cross section
are due to the $N(1535)$ and $\Delta(1600)$ resonances, which yield 0.011~$\mu$b/GeV and
0.0078~$\mu$b/GeV, respectively. Although these are negligible compared to the dominant contributions, the interference of these resonances with the dominant ones could contribute at most about $\pm$20\% in the ideal case, where the interference is either constructive or destructive throughout the whole phase space. We computed these interference terms numerically, and found that they are negligible compared to the $N(1520)$ and $N(1440)$ contributions. This remains true also when the uncertainties of the sub-leading contributions are taken into account.

Our calculation of $d \sigma/d M$ is consistent with the results of \cite{Zetenyi:2012hg} at $M=0.5$~GeV within the uncertainties discussed above. Moreover, the cross section in \cite{Zetenyi:2012hg} has been compared with preliminary HADES data and found to be in a reasonable agreement at $\sqrt{s}=1.49$ GeV and $M=0.5$~GeV \cite{HADES}. An alternative check of the reliability of the model is obtained by studying the process $\pi N  \rightarrow N\pi\pi$. We thus computed the neutral $\rho$ contribution of the differential cross section at 0.5~GeV invariant mass of the pion pair, including the $N(1520)$ resonance in both the  $s$- and $u$-channels. Taking the uncertainties discussed above into account, we obtain a value for $d \sigma/d M$ between 5 and 10~mb/GeV, which is consistent with the result of the partial wave analysis of the Bonn-Gatchina group, presented in \cite{HADES2}.

The double-differential cross section, $d\sigma/dM d\cos\theta_{\gamma^*}$ obtained from
$s$- and $u$-channel diagrams with the $N(1520)$ and $N(1440)$ resonances is shown in 
\fig\ref{fig:sigma_phys_dominant} as a function of the polar angle of the virtual photon, 
${\theta_{\gamma^*}}$. Here two of the curves correspond to the contributions of 
the two resonances without interference. In the other two, the interference terms are included, assuming either a positive or negative relative sign between the two resonance amplitudes.
In general, the interaction vertices of the resonances with pions and $\rho$ mesons can be complex, as a result of their microscopic structure \cite{Lutz:2001mi}.
This can result in an energy dependent relative phase of two resonance amplitudes between $0$ and $\pi$. Since this phase is unknown, we give the results for two limiting cases, assuming a positive or negative relative sign between the $N(1520)$ and $N(1440)$ amplitudes.

From Fig.~\ref{fig:sigma_phys_dominant} it is clear that the relative phase has a strong 
influence on the shape of the $\theta_{\gamma^*}$ dependence
of the differential cross section. Moreover, as discussed above the magnitude of the $N(1440)$ contribution is uncertain, which in turn affects the shape of the differential cross section. This suggests that the unknown phase and the coupling strength of the $N(1440)$ to the $N\rho$ channel can be constrained by data on the angular distribution. 

\begin{figure}[tb]
	\begin{center}
		\includegraphics{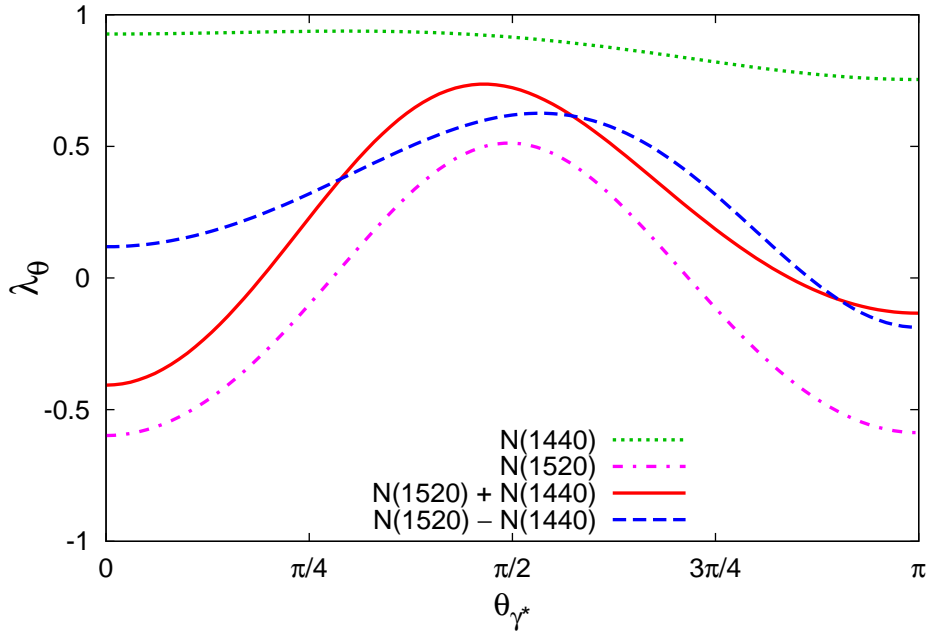}
		\caption{\label{fig:B_phys_dominant} The contribution of the two dominant resonances, $N(1440)$ and $N(1520)$,
			to the anisotropy coefficient, $\lambda_\theta$ at $\sqrt{s}=1.49$~GeV CM energy. 
			The various curves correspond to the same assumptions as in Fig.\ \ref{fig:sigma_phys_dominant}}.
	\end{center}
\end{figure}
\begin{figure}[bt]
	\begin{center}
		\includegraphics{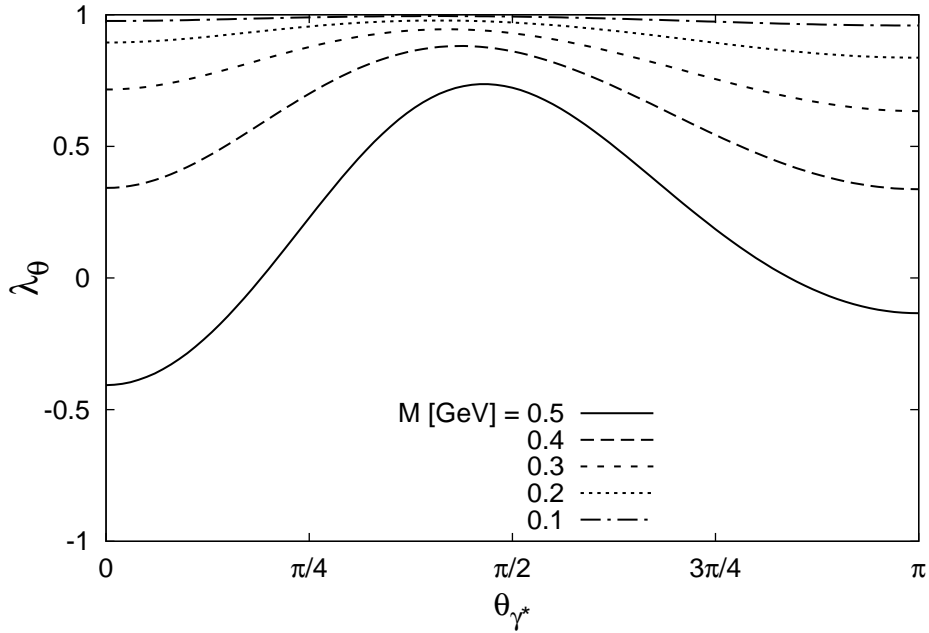}
		\caption{\label{fig:physical} The anisotropy coefficient $\lambda_\theta$ as a function
			of the virtual photon polar angle for various dilepton
			masses. The contributions of $s$- and $u$-channel diagrams of the
			two dominant resonances $N(1440)$ and $N(1520)$ and their interference term
			(with a positive relative sign) are included. The CM energy is $\sqrt{s}=1.49$~GeV.}
	\end{center}
\end{figure}

In \fig\ref{fig:B_phys_dominant} we show the dominant contributions to the anisotropy coefficient $\lambda_\theta$ 
as a function of ${\theta_{\gamma^*}}$. As in Fig.~\ref{fig:sigma_phys_dominant}, we show
results for the two limiting  assumptions for the relative phase of the two resonance amplitudes. In both cases, the shape of the curve approximately follows that of  the
$N(1520)$ contribution, which implies that it is only weakly affected by the uncertainties of the $N(1440)$ parameters. The anisotropy parameter $\lambda_\theta$ has a maximum around $\theta_{\gamma^*}=\pi/2$. Thus, virtual photons emitted 
perpendicular to the beam axis in the CM frame tend to be transversely polarized, while 
virtual photons emitted along the beam direction are almost unpolarized or, in the case of a 
positive relative phase between the two resonances, photons travelling in the forward direction tend 
to be longitudinally polarized.

Dileptons with a low invariant mass originate from the decay of a virtual photon which is
close to its mass shell. Such virtual photons must be predominantly transversely polarized. 
Consequently the $\lambda_\theta$ coefficient of the resulting dileptons is close to unity. This can be
seen in Fig. \ref{fig:physical}, where we show the $\theta_{\gamma^*}$ dependence of the $\lambda_\theta$
coefficient for various values of the dilepton invariant mass. These results include
the $s$- and $u$-channel diagrams of the dominant $N(1520)$ and $N(1440)$ resonances,
assuming a positive relative sign between the amplitudes.

The anisotropy coefficient $\lambda_\theta$ can be determined experimentally by employing \eq\eqref{eq:dsdm3}. Clearly this is very challenging, since such an analysis 
requires a triple-differential dilepton production cross section. For a fixed invariant mass $M$ 
and scattering angle $\theta_{\gamma^*}$, the $\lambda_\theta$ coefficient is obtained 
by extracting the dependence of the cross section on the electron angle $\cos^2 \theta_{e}$.
Nevertheless, the results shown in Fig.~\ref{fig:physical}
suggest that a rough binning both in $M$ and $\theta_{\gamma^*}$, e.g.\ $M > 0.3$ GeV and
three bins in $\theta_{\gamma^*}$, would be sufficient for extracting interesting information
on the polarization observable. 

\section{\label{sec:summary} Summary and outlook}

In this paper we studied the angular distribution of dileptons originating 
from the process $\pi N \rightarrow Ne^+e^-$ and presented numerical results
for the anisotropy coefficient $\lambda_\theta$ based on the assumption that the process
is dominated by intermediate baryon resonances. We employed effective 
Lagrangians to describe the interactions of baryon resonances with pions and 
photons. The coupling of the electromagnetic field to the 
baryon resonances is based on the vector meson dominance model.

The coupling constants of the model have been determined using information given by the Particle Data Group \cite{PDG}. Since the decay parameters of some of the baryon resonances are not very well known, our model contains uncertainties. However, the differential cross sections obtained using our model are in reasonable agreement with preliminary HADES data on dilepton production, and results on the neutral $\rho$ contribution extracted from a partial wave analysis of pion pair production. The shape of the anisotropy coefficient $\lambda_\theta$ as a function of the scattering angle is determined mainly by the $N(1520)$ resonance and hence it depends only weakly on the uncertainties of the model.

The anisotropy coefficient can in principle be determined in 
experiments by the HADES collaboration at GSI, Darmstadt. To this end, at least a rough 
binning of the triple-differential dilepton production cross section is 
needed. This requires high statistics, which is not easily achieved for such a rare 
probe. On the other hand, as we argued in this letter, the angular distributions provide
valuable additional information, which can help disentangle the various 
contributions to the dilepton production cross section and thus also provide novel 
information on the properties of baryon resonances. Consequently, high statistics data 
on pion induced dilepton production would provide important constraints on the elementary dilepton 
production mechanism as well as on the structure of baryons.

The calculation presented here is clearly exploratory. In future studies, several aspects 
of the model should be improved. First of all the model dependence 
of the predictions needs to be addressed. This can be done e.g.\ by repeating the 
calculation with different effective Lagrangians. A complementary approach, formulated in 
terms of helicity amplitudes or partial wave amplitudes, could provide
a systematic framework for exploring the various contributions to the scattering amplitude.

A previous study suggests that at the CM energy of the HADES experiment a major
part of the pion photoproduction cross section is probably due to non-resonant Born contributions 
\cite{Zetenyi:2012hg}. Consequently, these Born terms may significantly influence 
also the angular distributions of dilepton production and the anisotropy coefficient
in pion-nucleon collisions. Thus, their contribution to $\lambda_\theta$ should
be assessed.

It is also known that the standard vector meson dominance model does not provide
a satisfactory description of the electromagnetic interaction of baryon resonances.
This can be improved e.g.\ by relaxing the universal coupling assumption~\cite{Kroll:1967it} for the 
photon coupling to baryons, as discussed e.g. in Ref.~\cite{Zetenyi:2012hg}.

Additional constraints on the model are provided by pion-nucleon collisions with other final states. 
In particular the one-pion and two-pion final states are measured at HADES with much better
statistics than for the dilepton final state. Investigation of these final states
in the framework of the same model would provide an independent check and a 
possibility to put tighter constraints on some of the parameters of the model. Two-pion
production can also proceed via an intermediate $\rho$ meson, which makes this process
particularly interesting in the present context. 

\section*{Acknowledgments}
We wish to thank T.~Galatyuk, P.~Salabura, D.~Rischke and W.~Przygoda for valuable discussions. The work of E.S. was supported by VH-NG-823, Helmholtz Alliance HA216/EMMI and GSI. M.Z. was supported by the Hungarian OTKA Fund No. K109462, EMMI and HIC for FAIR. The work of B.F. was partially supported by EMMI.

\end{document}